**Defect Modulation Doping**

*Mirko Weidner, Anne Fuchs, Thorsten J.M. Bayer, Karsten Rachut, Getnet K. Deyu, Andreas Klein[*]*

Dr. Mirko Weidner
Technische Universität Darmstadt, Fachbereich Material- und Geowissenschaften, Fachgebiet Oberflächenforschung, Otto-Berndt-Str. 3, 64287 Darmstadt, Germany

Dr. Anne Fuchs
Technische Universität Darmstadt, Fachbereich Material- und Geowissenschaften, Fachgebiet Oberflächenforschung, Otto-Berndt-Str. 3, 64287 Darmstadt, Germany

Dr. Thorsten J.M. Bayer
Technische Universität Darmstadt, Fachbereich Material- und Geowissenschaften, Fachgebiet Oberflächenforschung, Otto-Berndt-Str. 3, 64287 Darmstadt, Germany

Karsten Rachut
Technische Universität Darmstadt, Fachbereich Material- und Geowissenschaften, Fachgebiet Oberflächenforschung, Otto-Berndt-Str. 3, 64287 Darmstadt, Germany

Getnet K. Deyu
Technische Universität Darmstadt, Fachbereich Material- und Geowissenschaften, Fachgebiet Oberflächenforschung, Otto-Berndt-Str. 3, 64287 Darmstadt, Germany

Prof. Dr. Andreas Klein
Technische Universität Darmstadt, Fachbereich Material- und Geowissenschaften, Fachgebiet Oberflächenforschung, Otto-Berndt-Str. 3, 64287 Darmstadt, Germany
E-mail: aklein@surface.tu-darmstadt.de

**Abstract:**
The doping of semiconductor materials is a fundamental part of modern technology, but the classical approaches have in many cases reached their limits both in regard to achievable charge carrier density, as well as mobility. Modulation doping, a mechanism that exploits the energy band alignment at an interface between two materials to induce free charge carriers in one of them, has been shown to circumvent the mobility restriction. Due to an alignment of doping limits by intrinsic defects, however, the carrier density limit cannot be lifted using this approach. Here we present a novel doping strategy using defects in a wide band gap material to dope the surface of a second semiconductor layer of dissimilar nature. We show that by depositing an insulator on a semiconductor material, the conductivity of the layer stack can be increased by seven orders of magnitude, without the necessity of high temperature processes or epitaxial growth. This approach has the potential to circumvent limits to both carrier mobility and density, opening up new possibilities in semiconductor device fabrication, particularly for the emerging field of oxide thin film electronics.

## 1. Introduction

The controlled doping of semiconductors is a prerequisite for modern technology. The amount of charge carriers that can be introduced into the host material by conventional substitutional doping is known to be restricted by material-specific doping limits, which are caused by the formation of compensating defects due to the Fermi level dependent defect formation energies.[1,2] Another drawback of the classical doping approach by elemental substitution is the congruent decrease of charge carrier mobility $\mu$ with increasing charge carrier density $n$, caused by ionized impurity scattering.[3] In the field of transparent oxide semiconductors, these physical limits have precluded a significant increase of achievable film conductivity for the past 25 years.[4,5]

By the invention of the modulation doping technique, in which the dopant impurities are spatially separated from the transport layer, it has become possible to circumvent the limitation of charge carrier mobility in highly doped semiconductors. This is achieved by forming an interface between one low doped and one highly doped semiconductor with different band gaps.[6,7] Electrical transport parallel to that interface can then take place on the low doped side of the interface, mostly unaffected by Coulomb scattering with the charged dopant impurities that supply the free charge carriers. The conductive interface can be described in terms of a two-dimensional electron gas (2DEG). The effect was first demonstrated for GaAs/Ga$_{1-x}$Al$_x$As interfaces by Störmer, Dingle et al.[8,9] and led to the discovery of the fractional quantum Hall effect.[10] More recently, interface doping effects have also been reported for ZnO/ZnMnO,[11] and ZnO/ZnMgO[12] structures.

While this 'conventional' modulation doping mechanism can alleviate the carrier mobility restriction in highly doped semiconductors, it cannot overcome the carrier density limit imposed by formation of compensating defects. This restriction is caused by the general use of two materials of similar chemical and lattice structure, such as GaAs and Ga$_{1-x}$Al$_x$As or ZnO and ZnMnO, for the formation of the doped interface. It will be shown in this contribution that using two chemically and structurally dissimilar materials instead, a careful control of interface properties can be used to induce previously unattainable charge carrier densities in one of them. The classical doping limits can be circumvented by this approach if the formation of compensating defects in the doped material is kinetically suppressed. The proposed doping mechanism is referred to as *defect modulation doping,* and is expected to contribute to future development of microelectronic and optoelectronic devices. It offers a novel approach to increase the electrical conductivity of a thin film stack. This can be achieved by tailoring an interface between two materials, which do not have to be conducting on their own (interface engineering). In contrast to classical modulation doping approaches, the process conditions are compatible with large area applications and low cost processes.[13] Furthermore, the materials chosen in this study are transparent, offering a wide range of application in the emerging field of transparent electronics.[14,15]

## 2. The concept of defect modulation doping

The Fermi energy, $E_F$, relative to a material's band structure defines the density of free charge carriers. In a semiconductor or insulator material, the number of free electrons or holes is given by the difference in energy of $E_F$ from conduction band minimum ($E_{CBM}$) and valence band maximum ($E_{VBM}$), respectively.

The maximum density of free electrons in a material, and therefore the highest achievable Fermi level position $E_{F,max}$, is limited by the formation of compensating acceptor-type defects, which will by nature trap free electrons and prohibit them from contributing to charge transport. The formation enthalpies of charged defects are a function of Fermi level position[16,17] limiting the range of the Fermi energy to values where the defect formation enthalpies are positive. The total accessible range of Fermi level

$$\Delta E_{F,acc} = E_{F,max} - E_{F,min} \qquad (1)$$

is dictated by the specific formation enthalpies of intrinsic defects, as shown in **Figure 1a**. The resulting doping limits are typically aligned for similar materials, i.e. materials with the same crystal structure and iconicity.[1, 18-21] While those doping limits are caused by intrinsic defects such as vacancies or interstitials, an alignment of defect energies is also observed for extrinsic defects such as hydrogen or transition metal impurities.[22-24] As is the case for the classical example of modulation doping, GaAs/Ga$_{1-x}$Al$_x$As, most technologically relevant semiconductor interfaces are formed between similar materials, in order to minimize lattice mismatch. The Fermi level can therefore not be increased above or lowered below the doping limit by conventional modulation doping.

The defect modulation doping approach introduced in this work uses two dissimilar materials to circumvent the alignment of doping limits. The use of dissimilar materials removes the constraint of aligned doping limits, as depicted in Figure 1b. By aligning two dissimilar materials it is therefore, in principle, possible to obtain Fermi levels outside the doping limits in the host material. Such a situation can, from a thermodynamic point of view, only be achieved if defects in the host material cannot form spontaneously when the Fermi energy is raised during deposition of a modulation layer. Low processing temperatures are therefore required to realize defect modulation doping, in order to suppress the formation of compensating intrinsic defects.

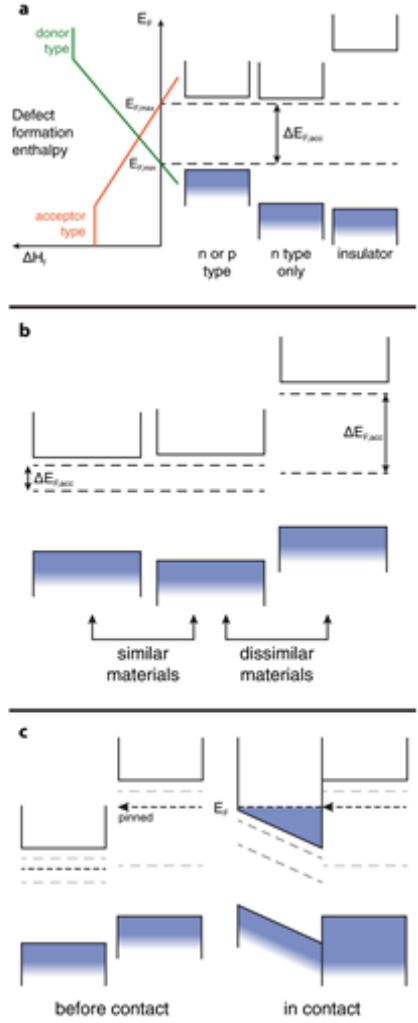

**Figure 1.** (a) The formation enthalpies of charge carrier compensating defects as a function of Fermi level position determine highest ($E_{F,max}$, given by acceptor-type defect formation) and lowest ($E_{F,min}$, dictated by donor-type defects) accessible Fermi level positions. The position of this range $\Delta E_{F,acc}$ relative to the band edges determine the dopability of a given material.
(b) Regardless of the band edge positions, the accessible Fermi level range is aligned on an absolute energy scale between similar materials, due to a similar defect chemistry that is largely determined by cation-anion binding energy and crystal structure. Between dissimilar materials, the accessible Fermi level ranges differ.
(c) Interface formation between two dissimilar materials, before contact (left) and in contact (right). The pinned Fermi level position on the right side of the interface forces the Fermi level above $E_{F,max}$ in proximity of the interface in the material on the left side. This induces the filling of conduction band states in the interface-near region, effectively doping one material beyond its classical doping limit. This is a direct representation of the defect modulation doping effect that is the topic of this study.

## 3. Results and Discussion

We demonstrate the effect of defect modulation doping by depositing a defective and amorphous insulator material ($Al_2O_3$)[25] on top of a polycrystalline wide band gap, transparent oxide semiconductor ($SnO_2$) in order to induce conduction electrons in the interface-near region of the latter. The modulation doping effect in $SnO_2$ (tetragonal structure, band gap 3.6 eV, formation enthalpy -577 kJ/mol[26]) is not achieved by the introduction of substitutional dopants, but instead by pinning of the Fermi level in the defective $Al_2O_3$, deposited at low process temperature, and at its interface to $SnO_2$. It has been shown previously that by using atomic layer deposition (ALD), defective $Al_2O_3$ can be reproducibly synthesized with the Fermi level pinned 4.5 eV above the valence band maximum.[27] As the valence bands of $SnO_2$ and $Al_2O_3$ are at similar energies,[28] it is expected that the Fermi energy is also 4.5 eV above the valence band maximum near the interface in $SnO_2$. This is higher than what can be achieved by conventional doping. The modulation doping is enabled by the low processing temperature of the ALD process, which prevents the formation of compensating defects.[29]

In situ X-ray photoelectron spectroscopy (XPS) measurements, comparing the $Sn3d_{5/2}$ emission line of an unintentionally doped film before and after deposition of an $Al_2O_3$ modulation layer, are shown in **Figure 2b**. $Al_2O_3$ deposition results in a binding energy shift from 486.33 eV to 487.17 eV. This shift is accompanied by peak broadening, i.e. an increase of emission line FWHM from 1.23 eV to 1.55 eV. Both effects are related to an increased Fermi level position in the sampled $SnO_2$ volume. More specifically, we can attribute the peak broadening in part by conduction-electron screening of the photohole,[30] as well as a downward band bending near the interface.[31] According to these effects the observed binding energy shift underestimates the actual change in Fermi level position at the interface.

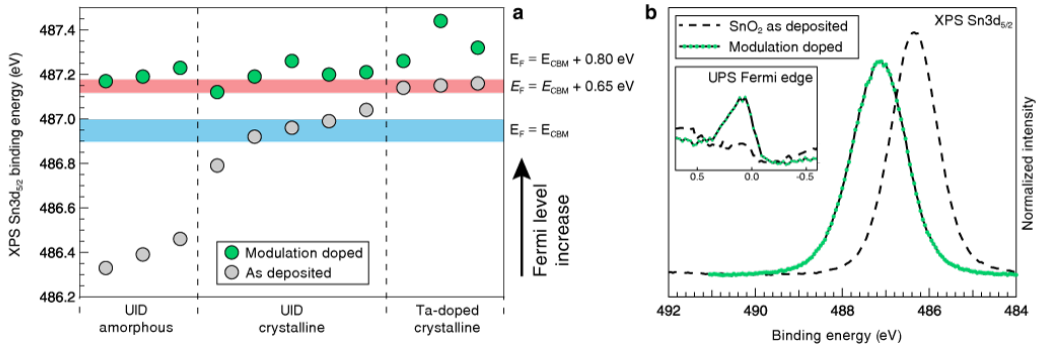

**Figure 2**: Comparison of $SnO_2$ tin oxide photoemission data before and after deposition of an $Al_2O_3$ modulation layer. Right side: the defect modulation doping results in an increased Fermi level position, reflected by a $Sn3d_{5/2}$ emission line shift, as well as the filling of conduction band states (inset). Left side: Comparison of $Sn3d_{5/2}$ binding energy values before (grey) and after (green) modulation doping. The blue bar indicates typical values of UID samples. The red bar indicates the highest binding energies achieved by conventional doping, which is surpassed by most modulation-doped samples. The increased $Sn3d_{5/2}$ binding energy is caused by a Fermi Level shift, but the correlation is not linear [33, 35].

The increase of the Fermi level position is furthermore verified by ultraviolet photoelectron spectroscopy (UPS), which is better suited than XPS for probing valence and conduction band emissions due to more favourable photoionisation cross sections. UPS is also the more surface sensitive of the two methods, allowing a more precise characterization of the interface-near region. Specifically, this makes it possible to visualise the filling of conduction band states upon modulation doping.[32] The filling of conduction band states is shown in the inset of Figure 2b. The two spectra closely resemble a comparison of unintentionally and of Sb-doped $SnO_2$,[30,33,34] indicating that conduction band states are indeed retroactively filled in the $SnO_2$ layer by the deposition of an $Al_2O_3$ layer.

Figure 2a compares respective binding energies of the Sn3d$_{5/2}$ emission line measured in-situ on tin dioxide surfaces before and after deposition of the 1 nm thick Al$_2$O$_3$ modulation layer. Fermi level positions approximated from those binding energies are given on the right hand side of the plot. Sn3d$_{5/2}$ binding energy values of 486.90 – 487.00 eV correlate with a Fermi level position close to the conduction band minimum, typically found in unintentionally doped crystalline samples. Higher binding energies are found when samples are doped with Sb or Ta reflecting the increased Fermi level position.[33,36] Once $E_F$ is above the conduction band minimum $E_{CBM}$, however, core-level binding energy shifts will underestimate the actual Fermi level shift due to the aforementioned photohole screening effects.[30,33] Photoemission data shows that the highest Sn3d$_{5/2}$ binding energy observed so far in substitutionally doped SnO$_2$ is 487.20 ± 0.05 eV, which relates to a Fermi level position at 0.65 eV above $E_{CBM}$.[36] Figure 2a illustrates a significant increase in Sn3d$_{5/2}$ binding energy after Al$_2$O$_3$ deposition for all samples, indicating that the Fermi level position can be pushed up to 0.8 eV or even higher into the conduction band.

Electrical measurements further confirm the defect modulation doping mechanism. As only the interface-near region is affected by the discussed approach, the effect is more pronounced in very thin films. Electrical measurements of three unintentionally doped samples are presented in **Table 1**. Samples were characterised as-deposited (subscript 'dep') and after modulation doping by deposition of an Al$_2$O$_3$ modulation layer (subscript 'md'). Electrical measurements were performed *in vacuo* (two point probe) in order to exclude the influence of an electron accumulation layer due to surface contamination,[26,32,37] as well as under ambient conditions (four point probe).

**Table 1:** Electrical measurement data for as deposited (subscript 'dep') and modulation doped (subscript 'md') samples. The first group of columns identifies the sample by ID, type of sputter target, substrate temperature during deposition, and film thickness. The second group compares in vacuo sample current of uncontaminated films at 10 V. The third group compares Hall effect measurements performed under ambient conditions.

| Sample | | | | *In vacuo* conductivity | | *Ex situ* Hall effect | | | | | |
|---|---|---|---|---|---|---|---|---|---|---|---|
| ID | Target | $T_{sub}$ | $t_{film}$ | $I_{dep}$ | $I_{md}$ | $\sigma_{dep}$ | $n_{dep}$ | $\mu_{dep}$ | $\sigma_{md}$ | $n_{md}$ | $\mu_{md}$ |
| | | (°C) | (nm) | (A) | (A) | (S/cm) | (cm$^{-3}$) | (cm$^2$V$^{-1}$s$^{-1}$) | (S/cm) | (cm$^{-3}$) | (cm$^2$V$^{-1}$s$^{-1}$) |
| I | SnO$_2$ | 600 | 20 | 1.67 × 10$^{-8}$ | 2.12 × 10$^{-4}$ | 9.0 | 2.9 × 10$^{19}$ | 1.8 | 262 | 1.0 × 10$^{20}$ | 16 |
| II | SnO$_2$ | 25 | 10 | 1 × 10$^{-12}$ | 1.94 × 10$^{-6}$ | 3 × 10$^{-4}$ | - | - | 0.1 | - | - |
| III | Sn | 25 | 10 | 2.55 × 10$^{-10}$ | 1.11 × 10$^{-3}$ | 7 × 10$^{-5}$ | - | - | 301 | 1.4 × 10$^{20}$ | 14 |

For all samples, the *in vacuo* sample current was increased by several orders of magnitude after modulation layer deposition. The largest effect was found for sample III, with an increase of more than six orders of magnitude. Hall effect measurements in air confirm the defect modulation doping effect, furthermore providing information about charge carrier density $n$ and mobility $\mu$ for samples with sufficient conductivity. The *ex situ* Hall effect conductivity increase found in sample III agrees with the *in vacuo* data. The lower increase of properties found in samples I and II when measured in air is attributed to the formation of an electron accumulation layer in air,[26,37] which already increases the conductivity of the as-deposited samples.

A mobility increase of one order of magnitude is found for sample I and is attributed to a decreased grain boundary barrier height[38] in the interface-near volume, due to the significantly increased Fermi level position in that region. Charge carrier densities greater than $10^{20}$ cm$^{-3}$ are achieved. While these values do not *per se* prove that the material has been doped beyond the values achievable by substitutional doping,[36,39-41] it has to be kept in mind that the employed measurement provides a value averaged across the film thickness and therefore must by nature severely underestimate the charge carrier density in the 2DEG at the interface.

Control experiments were performed in order to minimize the possibility of data misinterpretation, i.e. the observed increased Fermi level position and film conductivity being caused by an effect other than the proposed defect modulation doping mechanism. Among the effects excluded based on these experiments are a chemical reduction of the SnO$_2$ surface, an annealing effect during the Al$_2$O$_3$ deposition and an electron accumulation layer that is caused solely by sample exposure to water during the ALD process. The performed control experiments are discussed in more detail in the Supplementary Information.

The defect modulation doping approach makes use of the Fermi level pinning on the Al$_2$O$_3$ side of the interface, which was found to be located around 4.5 eV above the valence band maximum (VBM) regardless of the employed substrate.[27] This pinning is, however, specific to the deposition process. A comparison with nominally identical interfaces using sputter-deposited Al$_2$O$_3$ films has revealed no comparable pinning level.[27]

The pinned Fermi level in the ALD grown Al$_2$O$_3$ thin films could either be caused by hydrogen impurities,[22] which could be incorporated during the ALD process. Alternatively, native defects such as aluminium vacancies and interstitials can cause Fermi level pinning in Al$_2$O$_3$.[42] In the latter case, a variation of the oxygen chemical potential during Al$_2$O$_3$ deposition would allow for tuning of the Fermi level pinning position. Under strongly oxidizing conditions it could, in principle, also be used to achieve p-type modulation doping by lowering the Fermi level in the substrate.

A band diagram of a 20 nm thick unintentionally doped SnO$_2$ film, sputter-deposited in the presence of excess oxygen, before and after atomic layer deposition of a 1 nm Al$_2$O$_3$ modulation layer is shown in **Figure 3**.

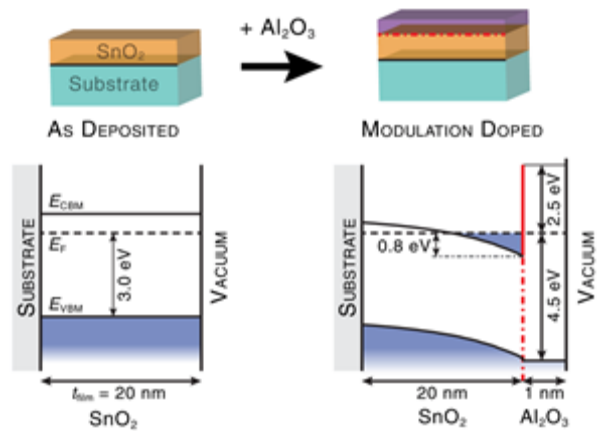

**Figure 3**: Band diagram of a 20 nm SnO$_2$ film before (left) and after (right) deposition of a 1 nm Al$_2$O$_3$ modulation layer. The SnO$_2$ tin oxide Fermi level is forced into the conduction band at the doped interface, but remains low near the interface to the substrate. At a position of 0.8 eV or more above the conduction band minimum, the Fermi level is situated above the classical doping limit (0.65 eV), demonstrating the success of the defect modulation doping approach.

This figure illustrates the specific result of this study on the defect modulation doping mechanism, which has been visualised in a generalized form in Figure 1c. Respective band gaps are 3.6 eV for $SnO_2$[43] and 7.0 eV for ALD-$Al_2O_3$.[44] After modulation doping, the precise Fermi level at the $SnO_2$ interface is unknown. Based on photoelectron spectroscopy data, it can conservatively be estimated to be 0.8 eV above the $SnO_2$ conduction band minimum at the interface with the $Al_2O_3$ modulation layer, but might very well be higher. This doping level is above the substitutional doping limit and is in good agreement with an $Al_2O_3$ Fermi level pinning at 4.5 eV above the valence band maximum and a negligible valence band offset at the $SnO_2/Al_2O_3$ interface.[17,28]

## 4. Summary and Conclusion

In summary, the concept of defect modulation doping has been proposed and demonstrated in this work using photoelectron spectroscopy and electrical measurements on $SnO_2$ thin films with ALD grown $Al_2O_3$ doping layers. In contrast to conventional modulation doping, the pinned Fermi level position of a defective insulator material is used to dope the interface-near region of a dissimilar semiconductor material. In principle, no substitutional dopant elements are required for this doping process. The presented data indicates that defect modulation doping can be used to circumvent traditional doping limits, additionally to increasing the carrier mobility by the spatial separation of dopants and transport channel, as in conventional modulation doping. Overcoming the doping limit is particularly facilitated by the possibility to use low processing temperatures. The approach also completely lifts the constraint regarding the crystalline structures and lattice constants of the materials combination. This fact, combined with the low cost of the employed materials, results in considerable potential in regard to widespread application in consumer level electronic devices. Highly conductive thin film stacks, especially if optically transparent, lead to a significant decrease of Ohmic losses in optoelectronic devices such as displays, LEDs and solar cells, and may therefore increase device efficiency. Defect modulation doping might also contribute to induced two-dimensional electron gases at $SrTiO_3/Al_2O_3$ interfaces and for enhanced electrical properties in $ZnO/Ga_2O_3$ multilayers.[45-47] The doping mechanism in Aluminium doped Zinc Oxide, synthesized by ALD of ZnO with intermittent $Al_2O_3$ layers, could also be reasonably explained by the currently discussed findings. Defects also very likely contribute to the variety of phenomena observed at $SrTiO_3/LaAlO_3$ interface.[48,49]

The reported possibility to retroactively 'switch on' electronic transport in low-temperature deposited films presents a completely new approach to thin film doping, and might have major implications for the design and fabrication of optoelectronic devices, MOSFETs and transparent electronics. The combination of dissimilar materials, which could be suited for this approach is by nature manifold, compared to finding suited combinations of similar semiconductor materials. Applicability of this approach is in principle only limited by the possibility to control the Fermi Level position in the defective material, and producing an interface with a suitable band alignment to the doped material. The lack of need for epitaxial growth is another significant upside, compared to the classical modulation doping approach.

## 5. Experimental Section

All samples were deposited on 1x1 cm$^2$ fused silica substrates, which had been cleaned in an ultrasonic bath of acetone for 15 minutes. Tin oxide films were sputter-deposited from circular planar targets with 2 inch diameter and 3 mm thickness at 25 W sputtering power, at a pressure 5x10$^{-3}$ mbar. The target-substrate distance was 8.6 cm, resulting in a deposition rate of 5-10 nm/min, depending on the oxygen content in the process gas. For elevated temperature deposition, samples were heated during deposition by a halogen light bulb.

All samples discussed in detail in this work were deposited at elevated oxygen flow ratios, providing excess oxygen during film growth, which is known to suppress the unintentional doping mechanism that is intrinsic to the material. Samples I and II were deposited from a ceramic SnO$_2$ sputter target, the former at a substrate temperature $T_{sub}$ = 600 °C and the second at room temperature. Sample III was deposited from a metallic tin target at room temperature.

For atomic layer deposition of Aluminium oxide films water and trimethylamuminium (TMA) were used as precursors, in a continuously evacuated UHV chamber. During deposition, samples were heated to ~200 °C by a halogen light bulb. Precursor introduction pulse lengths were 80 ms for TMA and 150 ms for water. Each precursor pulse was followed by 300 s pumping time. More details about the experimental setup used for Al deposition have been published by Bayer et al. [27].

In vacuo conductivity was measured in two-point geometry inside the UHV characterization chamber, in order to exclude the influence of an electron accumulation layer due to surface contamination [37]. Due to the possible influence of contact resistance, in vacuo conductivity data is only discussed in terms of sample current for an externally applied voltage of 10 V. Ohmic behaviour was confirmed by altering the voltage across 3 orders of magnitude, and reversing its sign.

Hall effect measurements in van der Pauw geometry were performed in air, using a home-made electrical setup and a Lakeshore electromagnet. Magnetic field strength was 1.3 T. As-deposited data presented in Table 1 are based on control group samples without a modulation layer.

Photoemission spectroscopy was performed using a Physical Electronics PHI 5700 multitechnique surface analysis system, using a monochromatized AlKα source for XPS, and a Helium discharge lamp for UPS measurements. The instrumental resolution, as determined from a room temperature measurement of the silver Fermi edge, is 374 meV for XPS and 104 meV for UPS.

**Acknowledgements**

This work was supported by the State of Hessen (Germany) within the LOEWE center AdRIA (Adaptronics, Research, Innovation, Applications), the German Science Foundation (DFG) within the graduate school TICMO (Tunable Integrated Components for Microwaves and Optics) and by the European Union's Horizon 2020 research and innovation programme under the Marie Sklodowska-Curie grant agreement No 641640 (European Joint Doctorate for Multifunctional Materials, EJD-funmat).

**References**

[1] S.B. Zhang, S.-H. Wei, A. Zunger, 'Microscopic Origin of the Phenomenological Equilibrium "Doping Limit Rule" in n-Type III-V Semiconductors', *Physical Review Letters* **2000**, *84*, 1232-1235.

[2] W. Walukiewicz, 'Intrinsic limitations to the doping of wide-gap semiconductors', *Physica B* **2001**, *302-303*, 123-136.

[3] R.B. Dingle, 'Scattering of electrons and holes by charged donors and acceptors in semiconductors', *Philosophical Magazine* **1955**, *46*, 831-840.

[4] T. Minami, 'Transparent conducting oxide semiconductors for transparent electrodes', *Semiconductor Science and Technology* **2005**, *20*, S35-S44.

[5] K. Ellmer, (2012), 'Past Achievements and Future Challenges in the Development of Optically Transparent Electrodes', *Nature Photonics* **2012**, *6*, 808-816.

[6] R. People, J. C. Bean, D. V. Lang, A. M. Sergent, H. L. Störmer, K. W. Wecht, R. T. Lynch, K. Baldwin, 'Modulation doping in $Ge_xSi_{1-x}$/Si strained layer heterostructures', *Applied Physics Letters* **1984**, *45*, 1231-1233.

[7] J. J. Harris, J.A. Pals, R. Woltjer, 'Electronic Transport in Low-Dimensional Structures', *Reports on Progress in Physics* **1989**, *52*, 1217-1266.

[8] H. Störmer, R. Dingle, A. Gossard, W. Wiegmann, M. Sturge, 'Two-dimensional electron gas at a semiconductor-semiconductor interface ', *Solid State Commun.* **1979**, *29*, 705-709.

[9] R. Dingle, H. L. Störmer, A. C. Gossard, W. Wiegmann, 'Electron Mobilities in Modulation-doped Semiconductor Heterojunction Superlattices', *Appl. Phys. Lett.* **1978**, *33*, 665.

[10] D. C. Tsui, H. L. Störmer, A. C. Gossard, 'Two-dimensional magnetotransport in the extreme quantum limit', *Phys. Rev. Lett.* **1982**, *48*, 1559.

[11] T. Edahiro, N. Fujimura, T. Ito, 'Formation of two-dimensional electron gas and the magnetotransport behavior of ZnMnO/ZnO heterostructure', *Journal of Applied Physics* **2003**, *93*, 7673-7675.

[12] H. Tampo, H. Shibata, K. Matsubara, A. Yamada, P. Fons, S. Niki, M. Yamagata, H. Kanie, 'Two-dimensional electron gas in Zn polar ZnMgO/ZnO heterostructures grown by radical source molecular beam epitaxy', *Appl. Phys. Lett.* **2006**, 89, 132113.

[13] R. F. P. Martins, A. Ahnood, N. Correia, L. M. N. P. Pereira, R. Barros, P. M. C. B. Barquinha, R. Costa, I. M. M. Ferreira, A. Nathan, E. E. M. C. Fortunato, 'Recyclable, Flexible, Low-Power Oxide Electronics', *Adv. Funct. Mater.* **2013**, *23*, 2153.

[14] K. Nomura, H. Ohta, A. Takagi, T. Kamiya, M. Hirano, H. Hosono, 'Room-temperature fabrication of transparent flexible thin-film transistors using amorphous oxide semiconductors', *Nature* **2004**, *432*, 488.

[15] E. Fortunato, P. Barquinha, R. Martins, 'Oxide Semiconductor Thin-Film Transistors: A Review of Recent Advances', *Adv. Mater.* **2012**, *24* 2945.

[16] S. B. Zhang, J. E. Northrup, 'Chemical potential dependence of defect formation energies in GaAs: Application to Ga self-diffusion', *Phys. Rev. Lett.* **1991**, *67*, 2339.

[17] A. Klein, 'Interface Properties of Dielectric Oxides', *J. Am. Ceram. Soc.* **2016**, *99*, 369.

[18] S. B. Zhang, S.-H. Wei, A. Zunger, 'A phenomenological model for systemization and prediction of doping limits in II-VI and I-III-$VI_2$ compounds', *J. Appl. Phys.* **1998**, *83*, 3192.

[19] Y.-J. Zhao, C. Persson, S. Lany, A. Zunger, 'Why can $CuInSe_2$ be readily equilibrium-doped n-type but the wider-gap $CuGaSe_2$ cannot?', *Appl. Phys. Lett.* **2004**, *85*, 5860.
[20] J. Robertson, S. J. Clark, 'Limits to doping in oxides', *Phys. Rev. B* **2011**, *83*, 075205.
[21] R. Schafranek, S. Li, C. Chen, W. Wu, A. Klein, '$PbTiO_3$/$SrTiO_3$ interface: Energy band alignment and its relation to the limits of Fermi level variation', *Phys. Rev. B* **2011**, *84*, 045317.
[22] C. G. Van de Walle, J. Neugebauer, 'Universal alignment of hydrogen levels in semiconductors, insulators and solutions', *Nature* **2003**, *423*, 626.
[23] J. M. Langer, C. Delerue, M. Lannoo, H. Heinrich, 'Transition-metal impurities in semiconductors and heterojunction band lineups', *Phys. Rev. B* **1988**, *38*, 7723.
[24] A. Zunger, 'Composition dependence of deep impurity levels in alloys', *Phys. Rev. Lett.* **1985**, *54*, 849.
[25] V. Miikkulainen, M. Leskelä, M. Ritala, R. L. Puurunen, 'Crystallinity of inorganic films grown by atomic layer deposition: Overview and general trends', *J. Appl. Phys.* **2013**, *113*, 021301.
[26] M. Batzill, U. Diebold, 'The surface and materials science of tin oxide', *Progress in Surface Science* **2005**, *79*, 47.
[27] T. J. M. Bayer, A. Wachau, A. Fuchs, J. Deuermeier, A. Klein, 'Atomic layer deposition of $Al_2O_3$ onto Sn-doped $In_2O_3$: Absence of self-limited adsorption during initial growth by oxygen diffusion from the substrate and band offset modification by Fermi level pinning in $Al_2O_3$', *Chem. Mater.* **2012**, *24*, 4503.
[28] S. Li, F. Chen, R. Schafranek, T. J. M. Bayer, K. Rachut, A. Fuchs, S. Siol, M. Weidner, M. Hohmann, V. Pfeifer, J. Morasch, C. Ghinea, E. Arveux, R. Günzler, J. Gassmann, C. Körber, Y. Gassenbauer, F. Säuberlich, G. Venkata Rao, S. Payan, M. Maglione, C. Chirila, L. Pintilie, L. Jia, K. Ellmer, M. Naderer, K. Reichmann, U. Böttger, S. Schmelzer, R. C. Frunza, H. Uršič, B. Malič, W.-B. Wu, P. Erhart, A. Klein, 'Intrinsic energy band alignment of functional oxides', *phys. stat. sol. (rrl)* **2014**, *8*, 571.
[29] P. Ágoston, C. Körber, A. Klein, M. J. Puska, R. M. Nieminen, K. Albe, 'Limits for n-type doping in $In_2O_3$ and $SnO_2$: A theoretical approach by first-principles calculations using hybrid-functional methodology', *J. Appl. Phys.* **2010**, *108*, 053511
[30] R. G. Egdell, J. Rebane, T. J. Walker, D. S. L. Law, 'Competition between initial- and final-state effects in valence- and core-level x-ray photoemission of Sb-doped $SnO2$', *Phys. Rev. B* **1999**, *59*, 1792.
[31] N. Ohashi, H. Yoshikawa, Y. Yamashita, S. Ueda, J. Li, H. Okushi, K. Kobayashi, H. Haneda, 'Determination of Schottky barrier profile at Pt/$SrTiO3$:Nb junction by x-ray photoemission ', *Appl. Phys. Lett.* **2012**, *101*, 251911.
[32] K. H. L. Zhang, R. G. Egdell, F. Offi, S. Iacobucci, L. Petaccia, S. Gorovikov, P. D. C. King, 'Microscopic Origin of Electron Accumulation in $In2O3$', *Phys. Rev. Lett.* **2013**, *110*, 056803.
[33] R. G. Egdell, T. J. Walker, G. Beamson, 'The screening response of a dilute electron gas in core level photoemission from Sb-doped $SnO_2$', *J. Electron Spectrosc.* **2003**, *128*, 59.
[34] P. A. Cox, R. G. Egdell, C. Harding, W. R. Patterson, P. J. Taverner, 'Surface properties of antimony doped tin(IV) oxide: a study by electron spectroscopy', *Surf. Sci.* **1982**, *123*, 179.

[35] C. Körber, P. Ágoston, A. Klein, 'Surface and Bulk Properties of Sputter Deposited Intrinsic and Doped $SnO_2$ Thin Films', *Sens. Actuators B* **2009**, *139*, 665.
[36] M. Weidner, J. Brötz, A. Klein, 'Sputter-deposited polycrystalline tantalum-doped SnO2 layers', *Thin Solid Films* **2014**, *555*, 173.
[37] H.-J. Michel, H. Leiste, K. D. Schierbaum, J. Halbritter, 'Adsorbates and their effects on gas sensing properties of sputtered SnO2 films', *Appl. Surf. Sci.* **1998**, *126*, 57.
[38] M. W. J. Prins, K.-O. Grosse-Holz, J. F. M. Cillessen, L. F. Feiner, 'Grain-boundary-limited transport in semiconducting $SnO_2$ thin films: Model and experiments', *J. Appl. Phys.* **1998**, *83*, 888.
[39] B. Stjerna, E. Olsson, C. G. Granqvist, 'Optical and electrical properties of radio frequency sputtered tin oxide films doped with oxygen vacancies, F, Sb, or Mo', *J. Appl. Phys.* **1994**, *76*, 3797.
[40] S. W. Lee, Y.-W. Kim, H. Chen, 'Electrical properties of Ta-doped $SnO_2$ thin films prepared by the metal–organic chemical-vapor deposition method ', *Appl. Phys. Lett.* **2001**, *78*, 350.
[41] M. Weidner, J. Jia, Y. Shigesato, A. Klein, 'Comparative study of sputter-deposited $SnO_2$ films doped with antimony or tantalum', *Physica Status Solidi (b)* **2016**, *153*, 923.
[42] J. R. Weber, A. Janotti, C. G. Van de Walle, 'Native defects in $Al_2O_3$ and their impact on III-V/$Al_2O_3$ metal-oxide-semiconductor-based devices ', *J. Appl. Phys.* **2011**, *109*, 033715.
[43] A. Schleife, J. B. Varley, F. Fuchs, C. Rödl, F. Bechstedt, P. Rinke, A. Janotti, C. G. Van de Walle, 'Tin dioxide from first principles: Quasiparticle electronic states and optical properties', *Phys. Rev. B* **2011**, *83*, 035116.
[44] K. Y. Gao, T. Seyller, L. Ley, F. Ciobanu, G. Pensl, A. Tadich, J. D. Riley, R. G. C. Leckey, '$Al_2O_3$ prepared by atomic layer deposition as gate dielectric on 6H-SiC(0001)', *Appl. Phys. Lett.* **2003**, *83*, 1830.
[45] S. W. Lee, Y. Liu, J. Heo, R. G. Gordon, 'Creation and Control of Two-Dimensional Electron Gas Using Al-Based Amorphous Oxides/SrTiO3 Heterostructures Grown by Atomic Layer Deposition', *Nano Letters* **2012**, *12*, 4775.
[46] Y. Z. Chen, N. Bovet, F. Trier, D. V. Christensen, F. M. Qu, N. H. Andersen, T. Kasama, W. Zhang, R. Giraud, J. Dufouleur, T. S. Jespersen, J. R. Sun, A. Smith, J. Nygård, L. Lu, B. Büchner, B. G. Shen, S. Linderoth, N. Pryds, 'A high-mobility two-dimensional electron gas at the spinel/perovskite interface of c-Al2O3/SrTiO3', *Nature Commun.* **2013**, *4*, 1371.
[47] Y.-H. Lin, H. Faber, J. G. Labram, E. Stratakis, L. Sygellou, E. Kymakis, N. A. Hastas, R. Li, K. Zhao, A. Amassian, N. D. Treat, M. McLachlan, T. D. Anthopoulos, 'High Electron Mobility Thin-Film Transistors Based on Solution-Processed Semiconducting Metal Oxide Heterojunctions and Quasi-Superlattices', *Advanced Science* **2015**, *2*, 1500058.
[48] A. Ohtomo, H. Y. Hwang, 'A high-mobility electron gas at the LaAlO3/SrTiO3 heterointerface', *Nature* **2004**, *427*, 423.
[49] L. Yu, A. Zunger, 'A polarity-induced defect mechanism for conductivity and magnetism at polar–nonpolar oxide interfaces', *Nature Commun.* **2014**, *5*, 5118.

Supporting Information

**Defect Modulation Doping**

*Mirko Weidner, Anne Fuchs, Thorsten J.M. Bayer, Karsten Rachut, Getnet K. Deyu, Andreas Klein*[*]

A number of experiments were performed in order to exclude all alternative explanations of the observed modification of $SnO_2$ films upon deposition of a modulation layer, namely an increased electrical conductivity and a much higher Fermi level position.

One alternative explanation would be a chemical reduction of the $SnO_2$ surface in the process of $Al_2O_3$ deposition, leading to an intrinsic doping of tin oxide due to oxygen deficiency. This theory is strongly contradicted by a set of experiments using extrinsically doped $SnO_2$ films. These films have a much higher carrier density and bulk Fermi level position than achievable by intrinsic doping mechanisms, yet the photoemission characterization of these films shows the same effect of a strongly increased surface Fermi level position after deposition of $Al_2O_3$.

A reduction of the SnO2 surface is furthermore known to be clearly visible in XPS and UPS valence band spectra[1,2] and is not observed after deposition of $Al_2O_3$ on an oxygen-rich $SnO_2$ surface. Conversely, deposition of a modulation layer on an already strongly reduced $SnO_2$ surface leads to the expected increase in Fermi level position, contradicting the theory of a doping effect induced by oxygen deficiency.

In order to exclude the possibility of the atomic layer deposition process being in some way responsible for the observed doping mechanism, for example due to annealing, the exposure to water at elevated temperature, or the incorporation of carbohydrates at the interface, the stack geometry was reversed. In this experiment, $SnO_2$ was deposited on ALD $Al_2O_3$. The film conductivity in this case was a factor 15 larger than that of a control film deposited in the absence of $Al_2O_3$. This experiment strongly suggests that formation of a $SnO_2$-$Al_2O_3$ interface is indeed causing the doping mechanism. The relatively low increase in conductivity in this stack geometry is attributed to the detrimental influence of the sputter-deposition of oxides (caused by the formation of high energy negatively charged particles), damaging the doped interface.

A further control experiment separately deals with the influx of heat and water during atomic layer deposition of $Al_2O_3$. To this end, the valve for TMA introduction into the chamber remained closed. The $SnO_2$ film was still heated to the same temperature and exposed to water at the same intervals and for the same duration. This did not induce any change of sample conductivity in air, as opposed to the defect modulation doping approach.

Preliminary experiments of $Al_2O_3$ deposition on a $In_2O_3$ film indicated that the modulation doping approach should be transferable to other oxides, resulting in a significantly increased Fermi level position at the inferface. This also indicates that specific properties of $SnO_2$, such as the dual valency of the Sn cation, are not responsible for the observed effect.

[1] R.G. Egdell, J. Rebane, T.J. Walker, and D.S.L. Law, *Competition between initial- and final-state effects in valence- and core-level x-ray photoemission of Sb-doped SnO2*, Phys. Rev. B **59** (1999), 1792.
[2] M. Weidner, J. Brötz, A. Klein, *Sputter-deposited polycrystalline tantalum-doped SnO2 layers*, Thin Solid Films **2014**, *555*, 173.